# Vortex reconnection rate, and loop birth rate, for a random wavefield.

J.H.Hannay,
H. H. Wills Physics Laboratory, University of Bristol, Tyndall Avenue, Bristol BS8 1TL, UK.

*Abstract*
A time dependent, complex scalar wavefield in three dimensions contains curved zero lines, wave 'vortices', that move around.  From time to time pairs of these lines contact each other and 'reconnect' in a well studied manner, and at other times tiny loops of new line appear from nowhere (births) and grow, or the reverse, existing loops shrink and disappear (deaths).  These three types are known to be the only generic events.  Here the average rate of their occurrences per unit volume, R, B, and D, is calculated exactly for a Gaussian random wavefield that has isotropic, stationary statistics, arising from a superposition of an infinity of plane waves in different directions.  A simplifying 'axis fixing' technique is introduced to achieve this.  The resulting formulas are expressed in terms of the power spectrum of the ensemble of plane waves: $R=W\sqrt{K_4^3/(12\pi^4 K_2(K_4-K_2^2))}$, and $B=D=(R/2)-(W/2)\sqrt{9 K_2^3/(16\pi^4)}$ where W is the standard deviation of angular frequencies, and $K_2$ and $K_4$ are the second and fourth moments of a wavevector component (say the x one). Thus reconnections are always more common than births and deaths combined.  As an expository preliminary, the case of two dimensions, where the vortices are points, is studied and the average rate of pair creation (and likewise destruction) per unit area is calculated to be $W\sqrt{(K_4-K_2^2)/(4\pi^4)}$.

## 1. *Introduction*

A complex scalar wavefield $\psi=f+ig$ in three dimensions generically has curved surfaces on which $f=0$ and others on which $g=0$.  The intersections of these two types of null surface are curved nodal lines along which the complex wavefield is zero, 'vortex lines', so called because of the locally helicoidal surfaces of constant phase.  If the wavefield is time dependent these lines move around.  From time to time closed loops can shrink to zero size and vanish, they die, and conversely they can be born out of nothing.  The apt analogy is with successive geographic contour lines, respectively near the top of a hill and the bottom of a valley.  Also from time to time pairs of vortex lines can touch and 'reconnect'.  They do so in the same way, locally, as successive contour lines of a geographic saddle point change from one pair of hyperbolas to the complementary pair (Fig 1).  These three types of event, births deaths and reconnections are the only generic possibilities [Berry and Dennis 2001, 2007, O'Holleran et al 2006, 2009].  Other motions such as two vortex curves sweeping across each other 'inertly', are not generic, instead they happen by two reconnections one after the other [Berry and Dennis 2012].

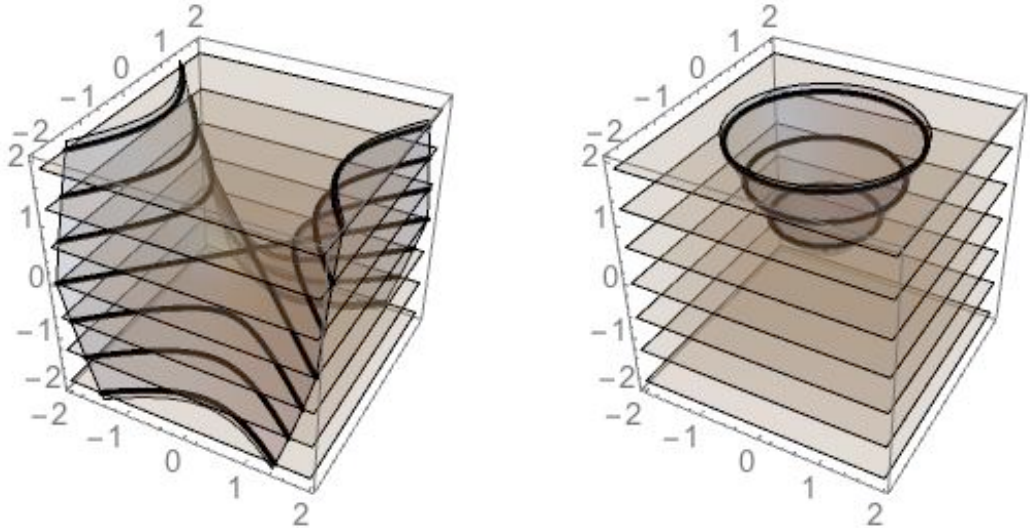

Fig 1.  Reconnection sequence (left), and loop birth sequence (right), with time increasing upwards, and the actual events at the origin.  For reconnection two vortices approach one another and join in a cross, and separate just like successive contours of a landscape saddle do.  For loop birth the vortices are like successive contours of a bowl shaped valley.  The analogy is this.  Contour lines are, by definition, the intersection of two surfaces:  the landscape surface and a horizontal plane.  Vortex lines are by definition the intersection of two curved surfaces in space: that of zero real part of the complex wavefunction, and that of zero imaginary part.  There is no loss of generality in displaying the sequence with one surface flat since it is only the (perpendicular) *separation* distance of the two surfaces that matters.  The time sequence is produced since the two surfaces are in relative motion – the plane moves upwards.  At the place and time of the event itself the two surfaces have their tangent planes coincident – in the pictures the horizontal plane through the origin is tangent to the landscape surface.

Here the average rate per unit volume of each of these three types of event is calculated exactly for a complex scalar function with stationary isotropic Gaussian randomness; a wavefunction $\psi$ arising from superposing (an infinite number of) plane waves $a\exp[i(k\mathbf{n}\cdot\mathbf{r}-\omega t+\varphi)]$.  Each wave has an independent uniformly random phase $\varphi$ ensuring stationary Gaussian randomness, and a uniformly random direction unit vector $\mathbf{n}$ ensuring isotropy.  The waves are free to have arbitrary real values of $a$ and $k$ (say positive without loss of generality), and of (real) angular frequency $\omega$.  Quite often in wave theory the angular frequency is taken as positive by definition (for example in the so called 'analytic signal' arising from complexifying a real valued wavefunction).  Here, although this is certainly allowed, it is not required, a mixture of plane waves with positive and negative frequencies is allowed too.

Usually, of course, plane waves obey a dispersion relation so that $\omega$ is a function of $k$.  An interesting special case is that when all the waves have the same speed $|\omega/k|=v$, thus obeying the standard wave equation $v^2\nabla^2\psi = \partial^2\psi/\partial t^2$  (non-dispersive waves). Another interesting case is that of Schrödinger quantum mechanics, $\omega \equiv k^2\hbar/2m$. Yet another special case, trivial for the present purposes, is when all the wave angular frequencies ω are equal; then the whole wavefield time dependence goes as exp(i$\omega$t) and the vortex lines do not move at all.

Previously many different statistics have been calculated for zeros, or equivalent features, in complex Gaussian random wavefields, for example [Berry and Dennis 2000, Dennis 2001, Berry and Hannay 1977, Taylor and Dennis 2015, Bogomolny and Schmit, 2007, Hannay 1996, 1998]. The averages to be evaluated here, though rather basic, have proved elusive [Berry and Dennis 2007]. Perhaps the main hurdle is formulating the required average without unmanageably many variables to be integrated over. Capitalizing on isotropy, an 'axis fixing' technique is used here to do this; even so there remains a 12-fold integral which, however, can be evaluated analytically. The resulting formulas (31) and (32) are simple despite the lengthy derivation. It may be that the technique, though quite elementary, could prove useful for other statistics in isotropic circumstances.

The formulas for the rates per unit volume of reconnections and tiny loop appearances and disappearances (births and deaths) ultimately depend on just a few (five, or ultimately four) 'moments' of the normalized power spectrum $\Phi(\omega,\mathbf{k})$ of the random superposition of plane waves, namely $\overline{k^2}$, $\overline{k^4}$, $\overline{\omega}$, $\overline{\omega^2}$, $\overline{\omega k^2}$, where the over-bar means the average, weighted by intensity, over all the plane waves – that is, integration with weighting $\Phi(\omega,\mathbf{k})$, over all wavevector space and all angular frequencies. But more directly, the formulas depend on a 16×16 matrix (mostly zeros) that is the inverse of a certain 'correlation matrix'. The matrix elements are proportional to one or other of the moments just listed, with simple rational number constants of proportionality given in appendix A. Thankfully the largest matrix (15) one needs to write down is 8×8 since the remaining 8 non-zero elements are self-correlations only (forming a purely diagonal 8×8 submatrix of the 16×16 matrix).

An easier quest than the main one is to seek the equivalent quantities for the case of just two space dimensions (and one time). Then the vortices are moving points not moving lines, and the generic events are the creation (or birth) of pairs, or the coalescence and disappearance (or death) of pairs. The average rates of these are obviously equal, and their calculation in the next section (2) serves as an instructive preparation for the main three dimensional calculation in section 3. Afterwards, analytical and numerical checks on the formulas in both two and three dimensions are described briefly in section 4.

Although it is often much neater for statistical calculations to use the complex variable ψ and its complex conjugate, rather than its real and imaginary parts ƒ and g, a feature of the main calculation here (the quartic in ƒ and g in an exponent in (17)) seems to spoil the neatness. So the separate real and imaginary parts ƒ and g are retained throughout.

*2. Random wavefield in two dimensions: births and deaths of vortex pairs.*

The vortices in two space dimensions are moving points, not moving lines. From time to time pairs of vortices appear from nowhere and separate (births), or the reverse, they annihilate and disappear (deaths). The average rate of births and deaths (twice either one) per unit area, $\Omega$, can be expressed, below, in a compact way (2), where $f(x,y,t)$ and $g(x,y,t)$ are the real and imaginary parts of the complex scalar wavefunction $\psi=f+ig$. Some introductory justification is in order. Partial derivatives of $f$ and $g$ will be denoted by subscripts.

At any one point and time, the average value of several different pairwise products of variables will be required. The variables are: $f$ or one of its partial derivatives, or $g$ or one of its partial derivatives. Examples would be $\langle f_{xx}f_t \rangle$ or $\langle f_y g_y \rangle$ or $\langle g_{xy}g \rangle$. Actually these particular examples are zero by isotropy or by phase randomness. The general rules are simple: the average of the pair product is zero if any of the following circumstances hold: the total number of $x$ subscripts is odd; the total number of $y$ subscripts is odd; the product is a mixed $fg$ type and the total number of all subscripts is even; the product is of $ff$ type and the total is odd; the product is of $gg$ type and the total is odd. The first two circumstances give zero by isotropy, and the remaining three give zero by phase randomness. Averages of products that are not forced to be zero by the above rules will typically be non-zero.

The statistical isotropy of $f$ and $g$, allows the formulation of an 'Axis fixing technique' which can considerably simplify averages. Consider any intrinsic, that is, coordinate rotation independent, quantity '•' depending on the function $f$, for example $(\nabla f)^2$, or $|\nabla f \times \nabla g|^2$, but not $f_x$ which depends on axis choice. Then the axis fixing asserts

$$\langle \bullet \rangle = \pi \langle \bullet \; \delta(f_x/f_y) \rangle \tag{1}$$

Here $f_x/f_y$ is tan of the angle $\theta$ of the contour of constant $f$ with the $x$ axis. But the quantity '•' does not depend on the choice of the $x$ axis. Therefore
$\langle \bullet \; \delta(f_x/f_y) \rangle = \langle \bullet \rangle \langle \delta(f_x/f_y) \rangle = \langle \bullet \rangle \langle \delta(tan\theta) \rangle = \langle \bullet \rangle \langle \delta(\theta) + \delta(\theta - \pi) \rangle = \langle \bullet \rangle / \pi$.

A simple example of the application of this axis-fixing technique would be to find an alternative way to calculate the average number of vortex points per unit area (this is an easy enough example to evaluate without axis fixing [Berry and Dennis 2000, Dennis 2001]). This average can be thought of as the average value of a new function which is a delta function sited on each vortex point in the plane (where $f=g=0$) and zero elsewhere. Each delta function must have the standard unit integral. Thus the starting expression for the average is $\langle \delta(f)\delta(g) |\nabla f \times \nabla g| \rangle$ where the Jacobian modulus factor ensures unit integral. Using the axis fixing technique this is replaced by $\pi \langle \delta(f)\delta(g) \; |\nabla f \times \nabla g| \; \delta(f_x/f_y) \rangle =$

$\pi \langle \delta(f)\ \delta(g)\ |f_y g_x|\ \delta(f_x/f_y) \rangle = \pi \langle \delta(f) \rangle \langle \delta(g) \rangle \langle f_y^2 \rangle \langle |g_x| \rangle \langle \delta(f_x) \rangle$, where in the last equality the identity $\delta(az)=|1/a|\delta(z)$ has been used. Evaluating the product of the five averages, the result is $\frac{1}{2\pi} \langle f_x^2 \rangle / \langle f^2 \rangle$, agreeing with the standard result.

The condition for a pair creation (birth) or annihilation (death) event is that the zero contour of *f* touches the zero contour of *g* tangentially. By isotropy, this mutual tangent direction is equally likely to be in any direction, and one is entitled to take it by definition as the *x* direction, invoking (1). One loses no generality thereby. The pair of vortex points approach each other, or recede from each other, along this direction in the limit. The starting formula for the average rate of births and deaths (twice either one) of pairs of vortices per unit area, which will be justified and then evaluated, is

$$\Omega = \pi \langle \delta(f)\ \delta(g)\ \delta(f_x)\ \delta(g_x)\ |f_y g_t - f_t g_y|\ |f_{xx} g_y - f_y g_{xx}| \rangle \qquad (2)$$

The justification of this expression is formally as follows – the intuition behind it is given in appendix B. The general local form of the functions *f* and *g* near the birth or death (where the curves *f*=0 and *g*=0 touch tangentially at the event *x'*=*y'*=*t*=0, *x'* and *y'* being the local tangent and normal coordinates), is $f = \frac{1}{2} F_{xx} x'^2 + F_y y' + F_t t$ and $g = \frac{1}{2} G_{xx} x'^2 + G_y y' + G_t t$ with all six coefficients being constants. The directions of the *x'*, *y'* axes are rotated from the directions of the *x*, *y* axes by a uniformly random angle *θ*. The statement justifying (2), not hard to verify, is that, with these temporary forms for *f* and *g*, *whatever* the values of the six constants,

$$\pi \int \int_0^{2\pi} \delta(f)\ \delta(g)\ \delta(f_x)\ \delta(g_x)\ |f_y g_t - f_t g_y|\ |f_{xx} g_y - f_y g_{xx}|\ dxdydtd\theta / 2\pi = 1 \qquad (3)$$

This ensures that each birth or death event counts with equal average weight, unity. The two modulus terms are essential – their interpretation is given in appendix B. (The limits on integrals are taken as ±∞ throughout unless otherwise specified, and the integrals may, as here, be multiple ones).

It remains to evaluate (2). All the derivatives that are unconstrained by delta functions need integrating over with their appropriate probability distributions (conditional on the constrained quantities). Also the standard integral representation [Berry and Dennis 2000] of a modulus quantity, $|\bullet| = \int (1 - \exp i\xi\bullet)/\pi\xi^2\ d\xi$, is to be used for each of the two modulus quantities in (2). (The exponential can be replaced by cos since sin is an odd function, and then the identity may be verified as the (indefinite) integral with respect to • of a better known one for the signum function: $\text{sgn}(\bullet) = \int \sin(\xi\bullet)/\pi\xi\ d\xi$ ).

At any given point in space-time there are ten variables whose pair product averages are involved in the calculation. From the rules stated above, of these ten, non-zero pair product averages only exist internally in the set $f, g_t, f_{xx}$ and in the set $g, f_t, g_{xx}$. The joint probability distribution for all ten relevant variables is

$$\frac{\gamma_{x,x}\gamma_{y,y}\sqrt{\det \gamma}}{(2\pi)^5} \exp[-\tfrac{1}{2}\gamma_{x,x}(f_x^2+g_x^2)] \exp[-\tfrac{1}{2}\gamma_{y,y}(f_y^2+g_y^2)] \times$$
$$\exp[-\tfrac{1}{2}(f,f_t,f_{xx},g,g_t,g_{xx})\bullet \gamma \bullet (f,f_t,f_{xx},g,g_t,g_{xx})]$$

(4)

where $\gamma_{x,x} = 1/\langle f_x^2 \rangle = 1/\langle g_x^2 \rangle = 1/\langle y_y^2 \rangle = 1/\langle g_y^2 \rangle = \gamma_{y,y}$, and $\gamma$ is

$$\gamma = \left( \begin{array}{ccc|ccc} \gamma_{0,0} & 0 & \gamma_{0,xx} & 0 & \gamma_{0,t} & 0 \\ 0 & \gamma_{t,t} & 0 & \gamma_{t,0} & 0 & \gamma_{t,xx} \\ \gamma_{xx,0} & 0 & \gamma_{xx,xx} & 0 & \gamma_{xx,t} & 0 \\ \hline 0 & \gamma_{t,0} & 0 & \gamma_{0,0} & 0 & \gamma_{0,xx} \\ \gamma_{0,t} & 0 & \gamma_{xx,t} & 0 & \gamma_{t,t} & 0 \\ 0 & \gamma_{t,xx} & 0 & \gamma_{xx,0} & 0 & \gamma_{xx,xx} \end{array} \right)$$

(5)

This matrix derives from its inverse: the correlation matrix (the inversion is supplied explicitly in the appendix A)

$$\gamma^{-1} = \left( \begin{array}{ccc|ccc} \langle f^2 \rangle & 0 & \langle ff_{xx} \rangle & 0 & \langle fg_t \rangle & 0 \\ 0 & \langle f_t^2 \rangle & 0 & \langle f_t g \rangle & 0 & \langle f_t g_{xx} \rangle \\ \langle f_{xx} f \rangle & 0 & \langle f_{xx}^2 \rangle & 0 & \langle f_{xx} g_t \rangle & 0 \\ \hline 0 & \langle gf_t \rangle & 0 & \langle g^2 \rangle & 0 & \langle gg_{xx} \rangle \\ \langle g_t f \rangle & 0 & \langle g_t f_{xx} \rangle & 0 & \langle g_t^2 \rangle & 0 \\ 0 & \langle g_{xx} f_t \rangle & 0 & \langle g_{xx} g \rangle & 0 & \langle g_{xx}^2 \rangle \end{array} \right)$$

(6)

The complementary patterns of zero entries in the 3×3 submatrices signifies that a permuting of the elements (from $f, f_t, f_{xx}, g, g_t, g_{xx}$ to $f, g_t, f_{xx}, g, f_t, g_{xx}$) would render the matrix block diagonal with two 3×3 blocks (and its inverse would have the same form). This permuting is not adopted (to keep the calculation parallel to that in three dimensions in the next section), but one can expect simplicity in the inverse matrix. Both the matrices $\gamma$ and $\gamma^{-1}$ are symmetric as a whole, and their off diagonal 3×3 submatrices are individually antisymmetric. One has $\langle fg_t \rangle = -\langle f_t g \rangle$ (the $fg_t$ is the sum of terms $-\omega\cos^2(k\mathbf{n}.\mathbf{r}-\omega t+\varphi)$, while $f_t g$ is the sum of $\omega\sin^2(k\mathbf{n}.\mathbf{r}-\omega t+\varphi)$, so averaged over the random phases they are equal and opposite). Likewise $\langle f_{xx} g_t \rangle = -\langle f_t g_{xx} \rangle$. Correspondingly it turns out that $\gamma_{0,t} = -\gamma_{t,0}$ and $\gamma_{xx,t} = -\gamma_{t,xx}$.

Then using the integral representation for the two modulus quantities as described, one has for the rate $\Omega$ of pair appearances plus disappearances per unit area:

$$\Omega = \pi \frac{\gamma_{x,x}\gamma_{y,y}\sqrt{\det\gamma}}{(2\pi)^5} \times$$

$$\int \exp[-\tfrac{1}{2}\gamma_{y,y}(f_y^2 + g_y^2)]\exp[-\tfrac{1}{2}(0,f_t,f_{xx},0,g_t,g_{xx})\bullet\gamma\bullet(0,f_t,f_{xx},0,g_t,g_{xx})]\times$$

$$\{1-\exp[i\mu'(f_y g_t - f_t g_y)]\}\{1-\exp[i\mu(f_{xx}g_y - f_y g_{xx})]\}\frac{d\mu'}{\pi\mu'^2}\frac{d\mu}{\pi\mu^2}df_y dg_y df_t dg_t df_{xx} dg_{xx}$$

(7)

A convenient way to implement the two unities in the braces is to introduce two two-term summations over 0 and 1

$$\Omega = \pi \frac{\gamma_{x,x}\gamma_{y,y}\sqrt{\det\gamma}}{(2\pi)^5} \sum_{\sigma=0}^{1}\sum_{\sigma'=0}^{1}(-1)^{\sigma+\sigma'} \times$$

$$\int \exp[-\tfrac{1}{2}\gamma_{y,y}(f_y^2 + g_y^2)]\exp[-\tfrac{1}{2}(0,f_t,f_{xx},0,g_t,g_{xx})\bullet\gamma\bullet(0,f_t,f_{xx},0,g_t,g_{xx})]\times$$

$$\exp[i\mu'\sigma'(f_y g_t - f_t g_y)]\exp[i\mu\sigma(f_{xx}g_y - f_y g_{xx})]\frac{d\mu'}{\pi\mu'^2}\frac{d\mu}{\pi\mu^2}df_y dg_y df_t dg_t df_{xx} dg_{xx}$$

(8)

Of course none of the four summed multiple integrals converges individually (regularized interpretation of the denominators would be required), but they do so together which is all that matters. The three exponentials depending on $f_y$ and $g_y$ can be integrated with respect to those variables yielding $(2\pi/\gamma_{y,y})\exp[-\tfrac{1}{2}(\mu\sigma f_{xx} - \mu'\sigma' f_t)^2/\gamma_{y,y} - \tfrac{1}{2}(\mu\sigma g_{xx} - \mu'\sigma' g_t)^2/\gamma_{y,y}]$. With temporary notation $\mu''' \equiv \mu'\sigma'/\sqrt{\gamma_{y,y}}$, $\mu'' \equiv \mu\sigma/\sqrt{\gamma_{y,y}}$ one has

$$\Omega = \pi\frac{\gamma_{x,x}\gamma_{y,y}\sqrt{\det\gamma}}{(2\pi)^5}\sum_{\sigma=0}^{1}\sum_{\sigma'=0}^{1}(-1)^{\sigma+\sigma'}\times\frac{2\pi}{\gamma_{y,y}}$$

$$\int \exp[-\tfrac{1}{2}\begin{pmatrix}f_t\\f_{xx}\\g_t\\g_{xx}\end{pmatrix}\begin{pmatrix}\gamma_{t,t}+\mu'''^2 & \mu'''\mu'' & 0 & \gamma_{t,xx}\\ \mu'''\mu'' & \gamma_{xx,xx}+\mu''^2 & \gamma_{xx,t} & 0\\ 0 & \gamma_{xx,t} & \gamma_{t,t}+\mu'''^2 & \mu'''\mu''\\ \gamma_{t,xx} & 0 & \mu'''\mu'' & \gamma_{xx,xx}+\mu''^2\end{pmatrix}\begin{pmatrix}f_t\\f_{xx}\\g_t\\g_{xx}\end{pmatrix}]\times \quad (9)$$

$$\frac{d\mu'}{\pi\mu'^2}\frac{d\mu}{\pi\mu^2}df_t dg_t df_{xx} dg_{xx}$$

(Sometimes, as here, row and column vectors are corrupted to the other for typographical convenience). The four integrals over $f_t, f_{xx}, g_t$ and $g_{xx}$ are Gaussians and can be done. Reverting to $\mu$ and $\mu'$,

$$\Omega = \pi \frac{\gamma_{x,x}\gamma_{y,y}\sqrt{\det\gamma}}{(2\pi)^5} \sum_{\sigma=0}^{1} \sum_{\sigma'=0}^{1} (-1)^{\sigma+\sigma'} \frac{2\pi}{\gamma_{y,y}} \times$$

$$\int \frac{(2\pi)^2}{\mu'^2 \sigma'^2 \gamma_{xx,xx} + \mu^2 \sigma^2 \gamma_{t,t} + \gamma_{t,t}\gamma_{xx,xx} - \gamma_{t,xx}^2} \frac{1}{\gamma_{y,y}} \frac{d\mu'}{\pi\mu'^2} \frac{d\mu}{\pi\mu^2} \qquad (10)$$

$$= \pi \frac{\gamma_{x,x}\gamma_{y,y}\sqrt{\det\gamma}}{(2\pi)^2} \frac{4\sqrt{\gamma_{t,t}\gamma_{xx,xx}}}{\pi\gamma_{y,y}^2(\gamma_{t,t}\gamma_{xx,xx} - \gamma_{t,xx}^2)^2} = \frac{1}{\pi^2}\sqrt{\overline{\omega^2} - \overline{\omega}^2}\sqrt{\overline{k_x^4} - \overline{k_x^2}^2}$$

This, then, is the formula for two dimensions. For the last equality use has been made of the elements of the **γ** matrix in terms of the correlation matrix (6), and hence the power spectrum set $\overline{k^2}$, $\overline{k^4}$, $\overline{\omega}$, $\overline{\omega^2}$, $\overline{\omega k^2}$, given in appendix A. A final step could optionally be taken: to express the formula in terms of the manifestly isotropic moments, using powers of $k$ instead of $k_x$. The conversions are: $\overline{k_x^2} = \overline{k^2}/2$ and $\overline{k_x^4} = 3\overline{k^4}/8$, as explained in appendix A. The birth and the death rate have to be equal (each to ½Ω) since the vortex population (per unit area) is steady. This comes about through the equal probabilities that, at the event, the relative curvature of $f$ and $g$ zero contours has the same or opposite sign to their relative velocity (the same sign giving a death, the opposite sign giving a birth). One may note that in the case of monochromatic waves, ω=constant, the rate of births and deaths is zero consistent with the vortex points being motionless (the time dependence of the wavefield exp(-iωt) does not affect the intensity – zeros stay zeros).

3. *Random wavefield in three dimensions: vortex births, deaths, and reconnections*

In three dimensions the 'contours' of zero value of $f$ are surfaces, as are those of $g$, and the events whose rates per unit volume are sought, are those when the two surfaces touch with a common tangent plane at the touching point. Whether the event is a reconnection of a pair of vortices, or the appearance (birth) or disappearance (death) of a vortex loop, depends on the relative curvature of the two surfaces – the local form of the gap between them at the moment of touching. If the separation of the surfaces as a function of position in the mutual tangent plane (a quadratic function, locally) is saddle-like then the event is a reconnection. Otherwise it is an appearance or disappearance (which of these depends on the relative velocity of the surfaces, as in two dimensions). This is easy to imagine if one of the two surfaces is locally a plane. If the other is saddle-like (negative Gaussian curvature), the intersection curves of the two surfaces as they move through each other before and after the event are hyperbolas, just like successive level contours near a geographical saddle. If the non-flat surface is bowl-like instead (positive Gaussian curvature) the successive contours are nested ellipses that grow from nothing at the event (or the reverse, shrink to nothing).

The axis fixing identity analogous to (1) for three dimensions is, for any intrinsic, that is, coordinate rotation independent, quantity • depending on the statistically isotropic function $f$:

$$\langle \bullet \rangle = 2\pi \left\langle \bullet \; \delta\!\left(\frac{f_x}{f_z}\right) \delta\!\left(\frac{f_y}{f_z}\right) \right\rangle \tag{11}$$

Here the requirement that $f_x/f_z$ and $f_y/f_z$ are zero fixes the direction of the gradient vector of $f$ to be parallel, or antiparallel to the z axis. But the quantity • does not depend on the choice of the z axis, so the average in (11) factorizes into the product $2\pi\langle\bullet\rangle\langle\delta(f_x/f_z)\delta(f_y/f_z)\rangle$. The latter average is $1/2\pi$ since the total solid angle $4\pi$ available for the isotropically directed gradient vector is halved to account for the two possible directions.

To use axis fixing for the events in question, births, deaths and reconnections, one can define the z axis to be that perpendicular to the mutual tangent plane of the touching surfaces $f=0$ and $g=0$ at the event. Then one has analogously to (2) the total rate per unit volume of reconnections, loop births and loop deaths:

$$\Omega_{B+D+R} = 2\pi \langle \delta(f)\,\delta(g)\,\delta(f_x)\,\delta(g_x)\,\delta(f_y)\,\delta(g_y) |f_z g_t - f_t g_z| \times \\ |(f_{xx}g_z - f_z g_{xx})(f_{yy}g_z - f_z g_{yy}) - (f_{xy}g_z - f_z g_{xy})(f_{yx}g_z - f_z g_{yx})| \rangle \tag{12}$$

Again, this will be justified and then evaluated. The evaluation is quite a long calculation, but once the expression has been reduced to a single remaining integration (27), the versions of it, (29) and (30), for the individual rates (per unit volume) for births, deaths and reconnections are easily extracted. Then the final integrations yield the formulas (31) and (32).

The justification of (12) would formally run as follows – the intuition behind it is given in appendix B. The general local form of the functions $f$ and $g$ near the birth, death or reconnection, where the surfaces $f=0$ and $g=0$ touch tangentially at the event $x'=y'=z'=t=0$, $x'$ and $y'$ being the local tangent coordinates and $z'$ the normal coordinate, is $f = \tfrac{1}{2}(F_{xx}x'^2 + 2F_{xy}x'y' + F_{yy}y'^2) + F_z z' + F_t t$ and $g = \tfrac{1}{2}(G_{xx}x'^2 + 2G_{xy}x'y' + G_{yy}y'^2) + G_z z' + G_t t$, with all ten coefficients being constants. (For definiteness, the $x', y'$ frame could be that making $F_{xy}-G_{xy}=0$ which diagonalizes the quadratic form representing the gap between the two surfaces, locally). The directions of the axes $x',y',z'$ are rigidly rotated with respect to the directions of the $x, y, z$ axes via Euler angles $\alpha, \beta, \gamma$ ($0<\alpha<2\pi, 0<\beta<\pi$, $0<\gamma<2\pi$), with all ten coefficients being constants. The quantities $\alpha$, $\cos\beta$, and $\gamma$ are each independently uniformly random. The statement justifying (12) is that, with these temporary forms for $f$ and $g$, whatever the values of the ten constants,

$$2\pi \int\!\!\int\!\int_0^{2\pi}\!\int_{-1}^{1}\!\int_0^{2\pi} \delta(f)\,\delta(g)\,\delta(f_x)\,\delta(g_x)\,\delta(f_y)\,\delta(g_y)|f_z g_t - f_t g_z| \times \\ |(f_{xx}g_z - f_z g_{xx})(f_{yy}g_z - f_z g_{yy}) - (f_{xy}g_z - f_z g_{xy})^2| \,dxdydzdtd\alpha d\gamma\,d(\cos\beta)/8\pi^2 = 1 \tag{13}$$

This ensures that each event, birth or death or reconnection, counts with equal average weight, unity. The two modulus terms are essential – their interpretation is given in appendix B.

Having set up the 3D expression (12) for the total rate per unit volume of births, deaths and reconnections, one can proceed analogously to (4) in two dimensions to evaluate (12). The same rules apply for the pair product zero averages, just with $z$ included. There is zero average if any of these apply: $x$ subscript total is odd; same for $y$; same for $z$; overall subscript total is even for a mixed $fg$ type; overall subscript total is odd for an $ff$ type; overall subscript total is odd for a $gg$ type. Thus, of the sixteen relevant variables below, non-zero correlations (pair product averages) exist only internally in the set $f, g_t, f_{xx}, f_{yy}$ and in the set $g, f_t, g_{xx}, g_{yy}$.

The joint probability distribution of the sixteen variables $f, f_x, f_y, f_z, f_t, f_{xx}, f_{xy}, f_{yy}, g, g_x, g_y, g_z, g_t, g_{xx}, g_{xy}, g_{yy}$ is

$$\frac{\Gamma_{x,x} \Gamma_{y,y} \Gamma_{z,z} \Gamma_{xy,xy} \sqrt{\det \Gamma}}{(2\pi)^8} \times$$

$$\exp\left[-\frac{1}{2}\left(\Gamma_{x,x}(f_x^2 + g_x^2) + \Gamma_{y,y}(f_y^2 + g_y^2) + \Gamma_{z,z}(f_z^2 + g_z^2) + \Gamma_{xy,xy}(f_{xy}^2 + g_{xy}^2)\right)\right] \times \quad (14)$$

$$\exp\left[-\frac{1}{2}(f, f_t, f_{xx}, f_{yy}, g, g_t, g_{xx}, g_{yy}) \bullet \Gamma \bullet (f, f_t, f_{xx}, f_{yy}, g, g_t, g_{xx}, g_{yy})\right]$$

where $\Gamma_{x,x} = 1/\langle f_x^2 \rangle = \Gamma_{y,y} = \Gamma_{z,z}$ by isotropy and

$$\Gamma = \begin{pmatrix} \Gamma_{0,0} & 0 & \Gamma_{0,xx} & \Gamma_{0,yy} & 0 & \Gamma_{0,t} & 0 & 0 \\ 0 & \Gamma_{t,t} & 0 & 0 & \Gamma_{t,0} & 0 & \Gamma_{t,xx} & \Gamma_{t,yy} \\ \Gamma_{xx,0} & 0 & \Gamma_{xx,xx} & \Gamma_{xx,yy} & 0 & \Gamma_{xx,t} & 0 & 0 \\ \Gamma_{yy,0} & 0 & \Gamma_{yy,xx} & \Gamma_{yy,yy} & 0 & \Gamma_{yy,t} & 0 & 0 \\ \hline 0 & \Gamma_{t,0} & 0 & 0 & \Gamma_{0,0} & 0 & \Gamma_{0,xx} & \Gamma_{0,yy} \\ \Gamma_{0,t} & 0 & \Gamma_{xx,t} & \Gamma_{yy,t} & 0 & \Gamma_{t,t} & 0 & 0 \\ 0 & \Gamma_{t,xx} & 0 & 0 & \Gamma_{xx,0} & 0 & \Gamma_{xx,xx} & \Gamma_{xx,yy} \\ 0 & \Gamma_{t,yy} & 0 & 0 & \Gamma_{yy,0} & 0 & \Gamma_{yy,xx} & \Gamma_{yy,yy} \end{pmatrix} \quad (15)$$

is the inverse of the correlation matrix (the inversion is made explicit in the appendix A). The average brackets apply to all elements:

$$\Gamma^{-1} = \left\langle \begin{pmatrix} ff & 0 & ff_{xx} & ff_{yy} & 0 & fg_t & 0 & 0 \\ 0 & f_t f_t & 0 & 0 & f_t g & 0 & f_t g_{xx} & f_t g_{yy} \\ f_{xx} f & 0 & f_{xx} f_{xx} & f_{xx} f_{yy} & 0 & f_{xx} g_t & 0 & 0 \\ f_{yy} f & 0 & f_{yy} f_{xx} & f_{yy} f_{yy} & 0 & f_{yy} g_t & 0 & 0 \\ \hline 0 & g f_t & 0 & 0 & gg & 0 & gg_{xx} & gg_{yy} \\ g_t f & 0 & g_t f_{xx} & g_t f_{yy} & 0 & g_t g_t & 0 & 0 \\ 0 & g_{xx} f_t & 0 & 0 & g_{xx} g & 0 & g_{xx} g_{xx} & g_{xx} g_{yy} \\ 0 & g_{yy} f_t & 0 & 0 & g_{yy} g & 0 & g_{yy} g_{xx} & g_{yy} g_{yy} \end{pmatrix} \right\rangle \quad (16)$$

As was the case in two dimensions (5), adjacent 4×4 submatrices have complementary patterns of zero entries signifying that permuting the set $f, f_t, f_{xx}, f_{yy}, g, g_t, g_{xx}, g_{yy}$ to $f, g_t, f_{xx}, f_{yy}, g, f_t, g_{xx}, g_{yy}$ would render the matrix block diagonal with two 4×4 blocks. Again this permuting is not adopted because the neatness would be only temporary, but the simplicity of the inverse matrix is to be expected. Both matrices as a whole are symmetric, and have duplicated 4×4 diagonal submatrix blocks (using the fact that from phase randomness $f$ and $g$ have the same statistics). Both have their two 4×4 off-diagonal submatrices individually antisymmetric, for example $\langle f_{xx} g_t \rangle = -\langle f_t g_{xx} \rangle$ and $\Gamma_{xx,t} = -\Gamma_{t,xx}$. Also the same pairs of elements are equal by isotropy (for example $\Gamma_{xx,xx} = \Gamma_{yy,yy}$).

There are various other relations between elements of the correlation matrix supplied in Appendix A. Indeed only six are independent (or five if standardized by normalization, as stated in the introduction). However the notation will be kept as it is for the moment so that the manipulations are easier to follow.

The sum of the rates (per unit volume) $\Omega_B$, $\Omega_D$ and $\Omega_R$ of births, deaths, and reconnections will be denoted $\Omega_{B+D+R}$ for short. The delta functions in the six variables $f, f_x, f_y, g, g_x, g_y$ mean that these can be integrated over straightaway and the two modulus expressions in (12) are represented by two auxiliary integrations over $\mu$ and $\mu'$ (just as in two dimensions).

$$\Omega_{B+D+R} = 2\pi \frac{\Gamma_{x,x}\Gamma_{y,y}\Gamma_{z,z}\Gamma_{xy,xy}\sqrt{\det\Gamma}}{(2\pi)^8} \times \int \exp\left[-\frac{1}{2}\Gamma_{xy,xy}(f_{xy}^2 + g_{xy}^2) - \frac{1}{2}\Gamma_{z,z}(f_z^2 + g_z^2)\right] \times$$

$$\exp\left[-\frac{1}{2}(0, f_t, f_{xx}, f_{yy}, 0, g_t, g_{xx}, g_{yy}) \bullet \Gamma \bullet (0, f_t, f_{xx}, f_{yy}, 0, g_t, g_{xx}, g_{yy})\right] \times$$

$$\left(1 - \exp[i\mu\{(f_z g_{xx} - f_{xx} g_z)(f_z g_{yy} - f_{yy} g_z) - (f_z g_{xy} - f_{xy} g_z)^2\}]\right) \times$$

$$\left(1 - \exp[i\mu'\{f_z g_t - f_t g_z\}]\right) \frac{d\mu'}{\pi\mu'^2} \frac{d\mu}{\pi\mu^2} df_z dg_z df_t dg_t df_{xx} dg_{xx} df_{yy} dg_{yy} df_{xy} dg_{xy}$$

(17)

Performing the $f_{xy}$, and $g_{xy}$ integrations, and simplifying to a 6×6 matrix by virtue of the two zero elements in the 'vector' in the $\Gamma$ exponential in (17):

$$\Omega_{B+D+R} = 2\pi \frac{\Gamma_{x,x}\Gamma_{y,y}\Gamma_{z,z}\Gamma_{xy,xy}\sqrt{\det\Gamma}}{(2\pi)^8} \times$$

$$\sum_{\sigma=0}^{1}(-1)^{\sigma}\int \exp\left[-\tfrac{1}{2}\Gamma_{z,z}(f_z^2+g_z^2)\right]\frac{2\pi}{\sqrt{\Gamma_{xy,xy}^2+i2\mu\sigma\Gamma_{xy,xy}(f_z^2+g_z^2)}} \times$$

$$\exp\left[-\frac{1}{2}(f_t,f_{xx},f_{yy},g_t,g_{xx},g_{yy})\begin{pmatrix}\Gamma_{t,t} & 0 & 0 & 0 & \Gamma_{t,xx} & \Gamma_{t,yy} \\ 0 & \Gamma_{xx,xx} & \Gamma_{xx,yy} & \Gamma_{xx,t} & 0 & 0 \\ 0 & \Gamma_{yy,xx} & \Gamma_{yy,yy} & \Gamma_{yy,t} & 0 & 0 \\ 0 & \Gamma_{xx,t} & \Gamma_{yy,t} & \Gamma_{t,t} & 0 & 0 \\ \Gamma_{t,xx} & 0 & 0 & 0 & \Gamma_{xx,xx} & \Gamma_{xx,yy} \\ \Gamma_{t,yy} & 0 & 0 & 0 & \Gamma_{yy,xx} & \Gamma_{yy,yy}\end{pmatrix}\begin{pmatrix}f_t \\ f_{xx} \\ f_{yy} \\ g_t \\ g_{xx} \\ g_{yy}\end{pmatrix}\right]\times$$

$$\exp[i\mu\sigma\{(f_z g_{xx}-f_{xx}g_z)(f_z g_{yy}-f_{yy}g_z)\}]\times$$

$$(1-\exp[i\mu'\{f_z g_t-f_t g_z\}])\frac{d\mu'}{\pi\mu'^2}\frac{d\mu}{\pi\mu^2}df_z dg_z df_t dg_t df_{xx} dg_{xx} df_{yy} dg_{yy}$$

(18)

The two-term σ sum here (σ=0 or 1), accounts for the unity associated with the μ expression in (17), the zero value of σ switching off the μ and giving the unity. The μ exponential can be incorporated into the matrix, and the μ' exponential is re-written in terms of the 'vector' $(f_t, f_{xx}, f_{yy}, g_t, g_{xx}, g_{yy})$:

$$\Omega_{B+D+R} = 2\pi \frac{\Gamma_{x,x}\Gamma_{y,y}\Gamma_{z,z}\Gamma_{xy,xy}\sqrt{\det\Gamma}}{(2\pi)^8} \times$$

$$\sum_{\sigma=0}^{1}(-1)^{\sigma}\int \exp\left[-\tfrac{1}{2}\Gamma_{z,z}(f_z^2+g_z^2)\right]\frac{2\pi}{\sqrt{\Gamma_{xy,xy}^2+i2\mu\sigma\Gamma_{xy,xy}(f_z^2+g_z^2)}} \times$$

$$\exp\left[-\frac{1}{2}(f_t,f_{xx},f_{yy},g_t,g_{xx},g_{yy})\cdot\Gamma'\cdot(f_t,f_{xx},f_{yy},g_t,g_{xx},g_{yy})\cdot\right]\times$$

$$(1-\exp[i\mu'(f_t,f_{xx},f_{yy},g_t,g_{xx},g_{yy})\cdot(-g_z,0,0,f_z,0,0)])\frac{d\mu'}{\pi\mu'^2}\frac{d\mu}{\pi\mu^2}df_z dg_z df_t dg_t df_{xx} dg_{xx} df_{yy} dg_{yy}$$

where

$$\Gamma' \equiv \left( \begin{array}{cccccc} \Gamma_{t,t} & 0 & 0 & 0 & \Gamma_{t,xx} & \Gamma_{t,yy} \\ 0 & \Gamma_{xx,xx} & \Gamma_{xx,yy} - i\mu\sigma g_z^2 & \Gamma_{xx,t} & 0 & i\mu\sigma f_z g_z \\ 0 & \Gamma_{yy,xx} - i\mu\sigma g_z^2 & \Gamma_{yy,yy} & \Gamma_{yy,t} & i\mu\sigma f_z g_z & 0 \\ \hline 0 & \Gamma_{xx,t} & \Gamma_{yy,t} & \Gamma_{t,t} & 0 & 0 \\ \Gamma_{t,xx} & 0 & i\mu\sigma f_z g_z & 0 & \Gamma_{xx,xx} & \Gamma_{xx,yy} - i\mu\sigma f_z^2 \\ \Gamma_{t,yy} & i\mu\sigma f_z g_z & 0 & 0 & \Gamma_{yy,xx} - i\mu\sigma f_z^2 & \Gamma_{yy,yy} \end{array} \right)$$

(19)

with

$$\det \Gamma' = D[-\Gamma_{t,t}(f_z^2 + g_z^2)(i\mu\sigma) + \Gamma_{t,t}(\Gamma_{xx,xx} + \Gamma_{xx,yy}) - 2\Gamma_{t,xx}^2] \times \\ [(f_z^2 + g_z^2)(i\mu\sigma) + (\Gamma_{xx,xx} - \Gamma_{xx,yy})]$$

(20)

where

$$D \equiv \left[\Gamma_{t,t}(\Gamma_{xx,xx} + \Gamma_{xx,yy}) - 2\Gamma_{t,xx}^2\right](\Gamma_{xx,xx} - \Gamma_{xx,yy})$$

(21)

Thus $D^2$ is the value of the determinant $\det\Gamma'$ when $\mu=0$, and this will be of use later (for (27)), when the ratio $\sqrt{\det\Gamma}/D$ is required; by Jacobi's theorem on determinants of submatrices, this ratio (squared) equals $1/\langle f^2 \rangle \langle g^2 \rangle = 1/\langle f^2 \rangle^2$, the reciprocal determinant of the submatrix of $\Gamma^{-1}$ complementary to $\Gamma'^{-1}$ (at $\mu=0$).

The integrals over $f_t, f_{xx}, f_{yy}, g_t, g_{xx}, g_{yy}$ and then the $\mu'$ integral can be done now:

$$\Omega_{B+D+R} = 2\pi \frac{\Gamma_{x,x}\Gamma_{y,y}\Gamma_{z,z}\Gamma_{xy,xy}\sqrt{\det\Gamma}}{(2\pi)^8} \times \\ \sum_{\sigma=0}^{1}(-1)^\sigma \int \exp\left[-\tfrac{1}{2}\Gamma_{z,z}(f_z^2+g_z^2)\right]\frac{2\pi}{\sqrt{\Gamma_{xy,xy}^2 + i2\mu\sigma\Gamma_{xy,xy}(f_z^2+g_z^2)}} \times \\ \sqrt{\frac{(2\pi)^6}{\det\Gamma'}}\left(1-\exp\left[-\tfrac{1}{2}\mu'^2(-g_z,0,0,f_z,0,0)\cdot(\Gamma')^{-1}\cdot(-g_z,0,0,f_z,0,0)\right]\right)\frac{d\mu'}{\pi\mu'^2}\frac{d\mu}{\pi\mu^2}df_z dg_z$$

(22)

$$= 2\pi \frac{\Gamma_{x,x}\Gamma_{y,y}\Gamma_{z,z}\Gamma_{xy,xy}\sqrt{\det\Gamma}}{(2\pi)^8} \times \\ \sum_{\sigma=0}^{1}(-1)^\sigma \int \exp\left[-\tfrac{1}{2}\Gamma_{z,z}(f_z^2+g_z^2)\right]\frac{2\pi}{\sqrt{\Gamma_{xy,xy}^2 + i2\mu\sigma\Gamma_{xy,xy}(f_z^2+g_z^2)}} \times \\ \sqrt{\frac{(2\pi)^6}{\det\Gamma'}}\sqrt{\frac{2}{\pi}}\sqrt{(-g_z,0,0,f_z,0,0)\cdot(\Gamma')^{-1}\cdot(-g_z,0,0,f_z,0,0)}\frac{d\mu}{\pi\mu^2}df_z dg_z$$

(23)

A couple of simplifications can usefully now be noted. One is that the matrix expression inside the square root turns out to be given by

$$(-g_z, 0, 0, f_z, 0, 0) \cdot (\Gamma')^{-1} \cdot (-g_z, 0, 0, f_z, 0, 0) = (f_z^2 + g_z^2) \frac{(\Gamma_{xx,xx} + \Gamma_{xx,yy})}{\Gamma_{t,t}\Gamma_{xx,xx} + \Gamma_{t,t}\Gamma_{xx,yy} - 2\Gamma_{t,xx}^2}$$

$$= (f_z^2 + g_z^2)(\Gamma_{xx,xx}^2 - \Gamma_{xx,yy}^2)/D$$

(24)

The other is that det$\Gamma'$ (20) is the product of two factors, each of which, when square rooted in (27), gives a branch point in the integrand. One of these two branch points coincides with that of $\sqrt{\Gamma_{xy,xy}^2 + i2\mu\sigma\Gamma_{xy,xy}(f_z^2 + g_z^2)}$; this follows because $\Gamma_{xx,xx} - \Gamma_{xx,yy} = \frac{1}{2}\Gamma_{xy,xy}$ since the left hand side equals $1/(\langle f_{xx}^2 \rangle - \langle f_{xy}^2 \rangle)$, and the right hand side equals $1/(2\langle f_{xy}^2 \rangle)$, and isotropy means that $\langle f_{xx}^2 \rangle = 3\langle f_{xy}^2 \rangle$ (see Appendix A). The consequence of this coincidence is that a pole results from the combination of these two coincident branch points.

Implementing these simplifications, and defining $r^2 = f_z^2 + g_z^2$ one has

$$\Omega_{B+D+R} =$$

$$2\pi \frac{\Gamma_{x,x}^3 \Gamma_{xy,xy} \sqrt{\det \Gamma}}{(2\pi)^8} \frac{2\pi}{\sqrt{2\Gamma_{xy,xy}}} \sqrt{\frac{2}{\pi}} \sqrt{(2\pi)^6} \frac{\sqrt{\Gamma_{xx,xx}^2 - \Gamma_{xx,yy}^2}}{\sqrt{D}} \sum_{\sigma=0}^{1} (-1)^\sigma \int \int_0^\infty \exp\left[-\tfrac{1}{2}\Gamma_{z,z} r^2\right] r \times$$

$$\frac{1}{\sqrt{D}\left[r^2(i\mu\sigma) + (\Gamma_{xx,xx} - \Gamma_{xx,yy})\right] \sqrt{-\Gamma_{t,t} r^2 (i\mu\sigma) + \Gamma_{t,t}(\Gamma_{xx,xx} + \Gamma_{xx,yy}) - 2\Gamma_{t,xx}^2}} \frac{d\mu}{\pi\mu^2} 2\pi r dr$$

(25)

Next define $v=\mu r^2$ so $(d\mu/\pi\mu^2)2\pi r dr$ becomes $(dv/\pi v^2)2\pi r^3 dr$. The $r$ integral can be done: $\int_0^\infty \exp[-\tfrac{1}{2}\Gamma_{z,z} r^2] r^4 dr = 3\sqrt{\pi/2\Gamma_{z,z}^5}$.

$$\Omega_{B+D+R} =$$

$$2\pi \frac{\Gamma_{x,x}^3 \Gamma_{xy,xy} \sqrt{\det \Gamma}}{(2\pi)^8 D} \frac{2\pi}{\sqrt{2\Gamma_{xy,xy}}} \sqrt{\frac{2}{\pi}} \sqrt{(2\pi)^6} \sqrt{\Gamma_{xx,xx}^2 - \Gamma_{xx,yy}^2} \, 3\sqrt{\pi/2\Gamma_{z,z}^5} \sum_{\sigma=0}^{1} (-1)^\sigma \times$$

$$\int \frac{1}{\left[(iv\sigma) + (\Gamma_{xx,xx} - \Gamma_{xx,yy})\right] \sqrt{-\Gamma_{t,t}(iv\sigma) + \Gamma_{t,t}(\Gamma_{xx,xx} + \Gamma_{xx,yy}) - 2\Gamma_{t,xx}^2}} 2\pi \frac{dv}{\pi v^2}$$

(26)

Making the two-term $\sigma$ sum explicit, this has the form

$$\Omega_{B+D+R} = Q \int \left(1 - \frac{1}{(1-v/ia)\sqrt{1-v/ib}}\right)\frac{dv}{v^2} \qquad (27)$$

where $a$, $b$, and the prefactor $Q$ are real constants depending on the set $\Gamma_{x,x}$, $\Gamma_{xx,xx}$, $\Gamma_{xx,yy}$, $\Gamma_{t,t}$, $\Gamma_{t,xx}$. Thus $a = \Gamma_{xx,xx} - \Gamma_{xx,yy}$, $b = -(\Gamma_{xx,xx} + \Gamma_{xx,yy} - 2\Gamma_{t,xx}^2/\Gamma_{t,t})$, and, using the remarks after (21) and (24),

$$Q = \frac{3\sqrt{\Gamma_{x,x}}\sqrt{\Gamma_{xx,xx}+\Gamma_{xx,yy}}}{4\pi^3 \langle f^2 \rangle \sqrt{\Gamma_{t,t}(\Gamma_{xx,xx}+\Gamma_{xx,yy})-2\Gamma_{t,xx}^2}} \qquad (28)$$

So for the integration in (27) there is a branch cut from $ib$ to infinity and a simple pole at $ia$, $a$ being positive and $b$ negative (Fig2). The origin is also a simple pole, not a double one because of the unity in the main bracket of (27). The integration contour (the real axis) passes straight through it and the 'principal part' is implied. This can be checked by examining the underlying circumstances. In the original introduction of $\mu$, the sign in front of it was chosen arbitrarily as positive. If it had been chosen negative instead the singularities would all have been reversed in sign. Taking half the sum of these two options eliminates the pole at the origin (at the expense of there being two contributions, in the integrand, not one). This 'half the sum' decomposition easily confirms the 'principal part' interpretation claimed.

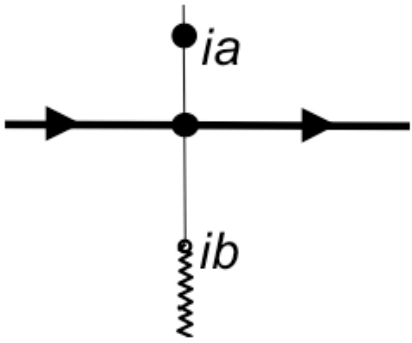

Fig 2. The complex $v$ plane for the contour integral (27) giving the summed average rates per unit volume of vortex births, deaths, and reconnections in a random wavefield. There is a simple pole at $ia$ and at the origin, and a branch point at $ib$, with a branch cut to $-i\infty$. If the contour of integration along the real axis is shifted slightly upwards, the integral gives twice the rate for reconnections alone. If shifted slightly downwards it gives the twice the combined rate for births and deaths. The shifted integrals are simply evaluated by pulling the contour fully upwards to infinity, picking up the residue at $ia$, and, for births and deaths, also the origin.

The integral in (27) can be directly evaluated by residues, but instead a natural split-up has merit. Being a principal part integral, (27) is the average of the two slightly displaced parallel contour integrals avoiding the origin, one just above the real axis and the other just below. These two turn out to be significant in their own right; the upper one is twice the rate per unit volume of reconnections $\Omega_R$ and the lower one is twice the combined rates per unit volume of births and deaths, $\Omega_B + \Omega_D$ ($=2\Omega_B=2\Omega_D$). The justification of these assertions is postponed to the end of this section, after presenting and discussing the evaluations. Writing $\varepsilon$ for an arbitrarily small positive real number (recall $b<0<a$):

$$\Omega_R = \frac{1}{2} Q \int_{-\infty+i\varepsilon}^{\infty+i\varepsilon} \left(1 - \frac{1}{(1-v/ia)\sqrt{1-v/ib}}\right) \frac{dv}{v^2} = \pi Q \frac{1}{a\sqrt{1-a/b}} \qquad (29)$$

and

$$\Omega_B + \Omega_D = \frac{1}{2} Q \int_{-\infty-i\varepsilon}^{\infty-i\varepsilon} \left(1 - \frac{1}{(1-v/ia)\sqrt{1-v/ib}}\right) \frac{dv}{v^2} = \Omega_R - \pi Q \left(\frac{1}{a} + \frac{1}{2b}\right) \qquad (30)$$

The second equality in (30) expresses the fact that the difference between the two oppositely shifted contour integrals comes from the residue at the origin.

Before justifying (29) and (30), they are now re-expressed in terms of moments of the wave power spectrum $\overline{k^2}$, $\overline{k^4}$, $\overline{\omega}$, $\overline{\omega^2}$, $\overline{\omega k^2}$, by using the substitutions from Appendix A (and the definitions of $a$, $b$ and $Q$ given in (28) above). The common prefactor $Q$ reduces to a simple form $Q = 3\sqrt{\overline{\omega^2} - \overline{\omega}^2} / (4\pi^3 \sqrt{\overline{k_x^2}})$. As in two dimensions one may note that this gives zero rates for monochromatic waves, $\omega$=constant, consistent with the vortices being motionless. The formulas could be expressed, to be manifestly isotropic, in terms of the magnitude $k$, rather than a component $k_x$, but they are neater as they stand. The conversions needed are $\overline{k_x^2} = \overline{k^2}/3$ and $\overline{k_x^4} = \overline{k^4}/5$ as explained in Appendix A. The final formulas, then, are these:

$$\Omega_R = \frac{1}{2\pi^2} \sqrt{\frac{\overline{k_x^4}^3 \left(\overline{\omega^2} - \overline{\omega}^2\right)}{3\overline{k_x^2}\left(\overline{k_x^4} - \overline{k_x^2}^2\right)}} \qquad (31)$$

$$\Omega_B = \Omega_D = \frac{1}{2}\Omega_R - \frac{3}{8\pi^2}\sqrt{\overline{k_x^2}^3 \left(\overline{\omega^2} - \overline{\omega}^2\right)} \qquad (32)$$

It is striking that these rates are both simply proportional to the standard deviation of angular frequencies, and that the moment $\overline{\omega k^2}$ is, in the end, not involved, so the four moments $\overline{k^2}$, $\overline{k^4}$, $\overline{\omega}$, $\overline{\omega^2}$ actually suffice.

An example is that of a thermal (black-body, Planck) power spectrum $\propto \omega k^2 / [\exp(\omega/W) - 1]\, dk$ for temperature $T$ where $W \equiv k_{Boltzmann} T / \hbar$. One then has to specify the dispersion relation $\omega(k)$ of the waves. For example with equal speeds $\omega=ck$, the constants $\overline{k_x^2}$, $\overline{k_x^4}$, $\overline{\omega}$, $\overline{\omega^2}$ are respectively:
$(20/3)(W/c)^2 \zeta(6)/\zeta(4)$, $168(W/c)^4 \zeta(8)/\zeta(4)$, $4W\zeta(5)/\zeta(4)$, $20W^2\zeta(6)/\zeta(4)$ .

With these one obtains, from (31) and (32), $\Omega_R = 4.2702 W^4/c^3$ and $\Omega_B + \Omega_D = 1.8526 W^4/c^3$.

Here, as always from (32), reconnections are more common than births and deaths combined, though for large kurtosis, $\overline{k_x^4}/\overline{k_x^2}^2$, of the power spectrum (note $\overline{k_x} = 0$), the ratio (births+deaths)/reconnections tends to unity as an upper limit. This ratio is also bounded below by $1 - 5/3\sqrt{3} \approx 0.03775$, realized when the kurtosis is least. From an inequality in appendix A, one has $\overline{k_x^4}/\overline{k_x^2}^2 \geq 9/5$, the equality limit being achieved when the wavelengths of the waves are all equal. In this limiting case the angular frequencies would also normally be equal to each other ('monochromatic') and therefore all rates would be zero.

As a technical, and possibly unphysical remark, if negative as well as positive angular frequencies are admitted equally for the constituent plane waves, the system has statistical time reversal symmetry. This means that the real and imaginary parts *f* and *g* of the wavefunction are completely independent random functions, and the whole analysis is considerably easier (for example $\overline{\omega} = \overline{\omega k^2} = 0$). In particular if the angular frequencies of monochromatic plane waves (equal wavelengths $2\pi/k$) were artificially liberated to be equal apart from a random sign: $\pm\omega$, then, with this 'monochromatic modulus' spectrum, as one might call it, the vortices move rather than being stationary. The rate per unit volume of reconnections is then $3\omega k^3/20\pi^2$ and the ratio (births+deaths)/reconnections is the limiting value 0.03775 stated above.

The justification of (29) and (30) is as follows. Since the contour in (27) passes directly through the pole at the origin, the 'principal part' is implied, and the sum of (29) and (30) correctly reproduces it. If (29) and (30) are correct, their difference $\Omega_B + \Omega_D - \Omega_R$ needs to be proved, now, to be the residue of the integrand of (27) at the origin $v=0$, times the constant $i\pi Q$. The feature distinguishing reconnections from births (tiny loop appearances) or deaths (tiny loop disappearances) is that the long expression inside the last modulus signs in (12),

$$(f_z g_{xx} - f_{xx} g_z)(f_z g_{yy} - f_{yy} g_z) - (f_{xy} g_z - f_z g_{xy})(f_{yx} g_z - f_z g_{yx}), \tag{33}$$

is negative instead of positive (that is, the determinant of the difference of the 2x2 curvature matrices of the touching zero surfaces of *f* and *g* is negative indicating a saddle form as in fig 1, left). Thus the quantity $\Omega_B + \Omega_D - \Omega_R$ is found by dropping the modulus signs on the long term in (12).

The integral representation for the modulus, that is, the $\mu$ integration in (17), $\int (......) d\mu/\pi\mu^2$, is no longer needed. Instead one may take the coefficient of $i\mu$ in (the Taylor expansion of) the (.....) part of the integrand (including of course the expression $\exp(i\mu...)$). This generates the required quantity (33) instead of its modulus. Equivalently this coefficient-taking can be postponed to (25), which has some of the other integrals inside the same (.....), already performed. The argument is, then, that the coefficient of $\mu$ in (…..), equals the coefficient of $1/\mu$ in (…..)$/\mu^2$, which amounts to the extraction of the residue. Converting to the variable $v$ this becomes the residue in the integrand of (27).

4. *Checks, analytical and numerical.*
Though the resulting formulas (31) and (32) are simple, the intermediate formulas in the derivation are rather involved, and checks, analytical and numerical, were very valuable for ensuring correctness, and noticing coincidences allowing progress to the simple formulas. These checks are described briefly here.

The two dimensional average rate of births and deaths $\Omega$ per unit area can be checked for a particular (rather peculiar) special wave dispersion relation $\omega(k_x, k_y)$, namely $k_x^2 + k_y^2 + (\omega/c)^2 = k^2$ with $c$ and $k$ constants, by using a trick. The idea is that these (2+1)D spacetime waves resemble a snapshot, say $t=0$, of a wavefunction of 3D monochromatic waves, $k_x^2 + k_y^2 + k_z^2 = k^2$, with the 'vertical' component $k_z$ of **k** replaced by $\omega/c$. Then the results from static vortex statistics in three dimensions [Berry and Dennis 2000] can be invoked. (Just as $k_z$ can be positive or negative, so can $\omega$; a rather artificial feature, physically, mentioned at the end of the previous section). A birth or a death in 2D corresponds to the 3D space curve having a horizontal tangent, that is, a local maximum or minimum in the z direction. The average number of such extremum points per unit volume can be calculated: it is the average length of vortex line per unit volume times the probability per unit length that a randomly chosen point on the line is an extremum. This latter means the following.

Mark points at equal, very small intervals $l$ along all vortex lines. A marked point is an extremum in $z$ (vertical) if the two short chords from it to its neighbours have opposite signs of their slopes, one rising, one falling (taking the three points in sequence). The magnitude of the difference in the slopes of the chords is the curvature $K$ of the vortex line times $l$, and the $z$ axis must therefore lie in an angular range $Kl$; one such range for a maximum, another for a minimum. The probability of the point being an extremum is therefore 2 $Kl$ /$2\pi$ (using isotropy). Therefore the required probability per unit length is $K/\pi$.

The average curvature was obtained by [Berry and Dennis 2000], as well as the average length of line per volume. For the monochromatic spectrum with all waves having wavenumber $k$, these averages are respectively $2k/\sqrt{15}$, and $k^2/3\pi$, giving a density of extrema of $2k^3/(3\pi^2\sqrt{15})$. With the speed $c=1$ this is the rate of births and deaths per unit area for the stated dispersion relation. There is agreement with (10) above; taking $k=1$, and $\langle f^2 \rangle = 1$ (this quantity cancels out in the answers, of course) one has the data
$\langle f^2 \rangle = 1$, $\langle f_t^2 \rangle = 1/3$, $\langle f_{xx}^2 \rangle = 1/5$, $\langle ff_{xx} \rangle = -1/3$. Together with (10) and (A1) to (A7), the agreement is confirmed.

In three dimensions various numerical checks can be made for the average rates per volume of births, deaths and reconnections. For instance, an intermediate integral, a 2-fold one, can be evaluated to quite a few digits of accuracy for any chosen spectrum of waves, and checked against the analytical formulas. Eqn (23) was used, with prior substitution to introduce the *r* variable whose integration can be performed analytically to reduce the 3fold integral to a 2-fold one. Or, for a cruder verification, the starting average expression (12), with the Gaussian probability distribution (14), can be evaluated approximately numerically by Monte-Carlo multidimensional Gaussian integration. Here the accuracy is, as would be expected, two or three digits for a million samples. For a more encompassing check, less accurate still, but not relying on the starting equation (12), a particular chosen realization of a random time-varying wavefield (with a chosen wave spectrum) can be inspected more or less automatically, and the events of births, deaths and reconnections simply counted within some spacetime 'box'. Such counts as have been made are consistent with (31) and (32).

*Acknowledgement*

I am grateful to John Nye for encouraging the use of specific examples for algebra checking/correction, particularly because, as he himself was finding in associated numerical searching and counting, reconnection events seemed to be strikingly common compared to births and deaths. I also thank Alexander Taylor for contributing a confirmatory numerical count by a separate means.

*Appendix A.  Relations between the constants (matrix elements).*

The explicit results of the inversion of the 6×6 $\gamma^{-1}$ correlation matrix (6) in two dimensions, and the 8×8 correlation matrix $\mathbf{\Gamma}^{-1}$ (16) in three dimensions are as follows. There are only six independent elements of the correlation matrix (constants), and these are taken as $\langle f^2 \rangle, \langle f_t^2 \rangle, \langle f_{xx}^2 \rangle, \langle fg_t \rangle, \langle ff_{xx} \rangle, \langle f_t g_{xx} \rangle$.
Commentary on the relations leading to the reduction to these six is given after the lists.

Two dimensions:

$$\gamma_{0,0} = \left( \langle f_t^2 \rangle \langle f_{xx}^2 \rangle - \langle f_t g_{xx} \rangle^2 \right) / \sqrt{\det \gamma^{-1}} \tag{A1}$$

$$\gamma_{t,t} = \left( \langle f^2 \rangle \langle f_{xx}^2 \rangle - \langle ff_{xx} \rangle^2 \right) / \sqrt{\det \gamma^{-1}} \tag{A2}$$

$$\gamma_{xx,xx} = \left( \langle f^2 \rangle \langle f_t^2 \rangle - \langle fg_t \rangle^2 \right) / \sqrt{\det \gamma^{-1}} \tag{A3}$$

$$\gamma_{0,xx} = \left( -\langle f_t^2 \rangle \langle ff_{xx} \rangle - \langle fg_t \rangle \langle f_t g_{xx} \rangle \right) / \sqrt{\det \gamma^{-1}} \tag{A4}$$

$$\gamma_{0,t} = \left( -\langle f_{xx}^2 \rangle \langle fg_t \rangle - \langle ff_{xx} \rangle \langle f_t g_{xx} \rangle \right) / \sqrt{\det \gamma^{-1}} \tag{A5}$$

$$\gamma_{t,xx} = \left( -\langle ff_{xx} \rangle \langle fg_t \rangle - \langle f^2 \rangle \langle f_t g_{xx} \rangle \right) / \sqrt{\det \gamma^{-1}} \tag{A6}$$

where

$$\sqrt{\det \gamma^{-1}} =$$
$$\langle f^2 \rangle \langle f_t^2 \rangle \langle f_{xx}^2 \rangle - \langle f_t^2 \rangle \langle ff_{xx} \rangle^2 - \langle f_{xx}^2 \rangle \langle fg_t \rangle^2 - \langle f^2 \rangle \langle f_t g_{xx} \rangle^2 - 2\langle ff_{xx} \rangle \langle fg_t \rangle \langle f_t g_{xx} \rangle$$
(A7)

Also $\gamma_{x,x} = -1/\langle ff_{xx} \rangle$.

Three dimensions:

$$\Gamma_{0,0} = \frac{4\langle f^2 \rangle^4 \langle f_{xx}^2 \rangle \left(2\langle f_t \rangle^2 \langle f_{xx} \rangle^2 - 3\langle f_t g_{xx} \rangle^2\right)}{9\sqrt{\det \Gamma^{-1}}}$$
(A8)

$$\Gamma_{0,xx} = \frac{6\langle f^2 \rangle^4 \langle f_{xx}^2 \rangle \left(\langle ff_{xx} \rangle \langle f_t^2 \rangle + \langle fg_t \rangle \langle f_t g_{xx} \rangle\right)}{9\sqrt{\det \Gamma^{-1}}}$$
(A9)

$$\Gamma_{0,t} = \frac{4\langle f^2 \rangle^4 \langle f_{xx}^2 \rangle \left(3\langle ff_{xx} \rangle \langle f_t g_{xx} \rangle + 2\langle fg_t \rangle \langle f_{xx}^2 \rangle\right)}{9\sqrt{\det \Gamma^{-1}}}$$
(A10)

$$\Gamma_{t,t} = \frac{4\langle f^2 \rangle^4 \langle f_{xx}^2 \rangle \left(3\langle ff_{xx} \rangle^2 - 2\langle f^2 \rangle \langle f_{xx}^2 \rangle\right)}{9\sqrt{\det \Gamma^{-1}}}$$
(A11)

$$\Gamma_{t,xx} = \frac{6\langle f^2 \rangle^4 \langle f_{xx}^2 \rangle \left(\langle fg_t \rangle \langle ff_{xx} \rangle + \langle f^2 \rangle \langle f_t g_{xx} \rangle\right)}{9\sqrt{\det \Gamma^{-1}}}$$
(A12)

$$\Gamma_{xx,xx} =$$
$$\frac{9\langle f^2 \rangle^4 \left(\langle f_t^2 \rangle \langle ff_{xx} \rangle^2 + 2\langle fg_t \rangle \langle ff_{xx} \rangle \langle f_t g_{xx} \rangle + \langle f^2 \rangle \langle f_t g_{xx} \rangle^2 + \langle f_{xx}^2 \rangle(\langle fg_t \rangle^2 - \langle f^2 \rangle \langle f_t^2 \rangle)\right)}{9\sqrt{\det \Gamma^{-1}}}$$
(A13)

$$\Gamma_{xx,yy} =$$
$$\frac{-3\langle f^2 \rangle^4 \left(3\langle f_t^2 \rangle \langle ff_{xx} \rangle^2 + 6\langle fg_t \rangle \langle ff_{xx} \rangle \langle f_t g_{xx} \rangle + 3\langle f^2 \rangle \langle f_t g_{xx} \rangle^2 + \langle f_{xx}^2 \rangle(\langle fg_t \rangle^2 - \langle f^2 \rangle \langle f_t^2 \rangle)\right)}{9\sqrt{\det \Gamma^{-1}}}$$
(A14)

where
$$\sqrt{\det \Gamma^{-1}} = \frac{2}{9}\langle f_{xx}^2 \rangle \times$$
$$\left(6\langle f_t^2 \rangle \langle ff_{xx} \rangle^2 + 12\langle fg_t \rangle \langle ff_{xx} \rangle \langle f_t g_{xx} \rangle + 6\langle f^2 \rangle \langle f_t g_{xx} \rangle^2 + 4\langle fg_t \rangle^2 \langle f_{xx}^2 \rangle - 4\langle f^2 \rangle \langle f_t^2 \rangle \langle f_{xx}^2 \rangle\right)$$
(A15)

Also $\Gamma_{xy,xy} = 1/\langle f_{xy}^2 \rangle = 1/\langle f_{xx} f_{yy} \rangle = 3/\langle f_{xx}^2 \rangle$ and $\Gamma_{x,x} = 1/\langle f_x^2 \rangle = -1/\langle ff_{xx} \rangle$. The set $\langle f^2 \rangle, \langle f_t^2 \rangle, \langle f_{xx}^2 \rangle, \langle fg_t \rangle, \langle ff_{xx} \rangle, \langle f_t g_{xx} \rangle$ are equal to $\langle f^2 \rangle$ times the set of moments of the normalized power spectrum $1, \overline{\omega^2}, \frac{1}{5}\overline{k^4}, -\overline{\omega}, -\frac{1}{3}\overline{k^2}, -\frac{1}{3}\overline{\omega k^2}$ in three dimensions, or $1, \overline{\omega^2}, \frac{3}{8}\overline{k^4}, -\overline{\omega}, -\frac{1}{2}\overline{k^2}, -\frac{1}{2}\overline{\omega k^2}$ in two dimensions.

Some of the relations between the pair product averages required above for the reduction to the six are general for any correlation function, namely $\langle f f_{xx} \rangle = -\langle f_x^2 \rangle$, and the same for $y$, and $\langle f_{xy}^2 \rangle = \langle f_{xx} f_{yy} \rangle$; the rule is that each transferred subscript between $f$s brings a minus sign. From phase randomness the same applies with $f$ replaced by $g$, corresponding averages being the same $\langle f f_{xx} \rangle = -\langle f_x^2 \rangle = \langle g g_{xx} \rangle = -\langle g_x^2 \rangle$, also crossed averages of $f$ and $g$ are related $\langle f_t g_{xx} \rangle = -\langle g_t f_{xx} \rangle$. The other relations derive from the isotropy and are mostly obvious: $\langle f_x^2 \rangle = \langle f_y^2 \rangle = \langle f_z^2 \rangle$ and $\langle f_{xx}^2 \rangle = \langle f_{yy}^2 \rangle$ (and the same with $f$ replaced by $g$), and $\langle f_t g_{xx} \rangle = \langle f_t g_{yy} \rangle$, and $\langle g_t f_{xx} \rangle = \langle g_t f_{yy} \rangle$. The only less trivial relation, explained below, is $\langle f_{xx} f_{yy} \rangle = \frac{1}{3} \langle f_{xx}^2 \rangle$ (and the same with $f$ replaced by $g$).

All the relations derive from their association with the moments of the normalized power spectrum, denoted by over-bars; $\langle f f_{x...xy...y} \rangle = (-1)^{(m+n)/2} \langle f^2 \rangle \overline{k_x^m k_y^n}$ where there are $m$ subscripts $x$ and $n$ subscripts $y$ (both even, or the average is zero). For a three dimensional spherically symmetric power spectrum one has
$\overline{k_x^m k_y^n} = \overline{k^{m+n}} \int \cos^{m+n}(\theta) \cos^m(\phi) \sin^n(\phi) \sin(\theta) d\theta d\phi / 4\pi$. For a two dimensional circularly symmetric power spectrum one has
$\overline{k_x^m k_y^n} = \overline{k^{m+n}} \int \cos^m(\phi) \sin^n(\phi) d\phi / 2\pi$. These angular integrals are in general given by the Beta function, but only those for low values of $m$ and $n$ are required. For $m=n=2$, one has $\overline{k_x^4} = 3\overline{k_x^2 k_y^2} (= \frac{1}{5}\overline{k^4})$ giving $\langle f_{xx} f_{yy} \rangle = \frac{1}{3} \langle f_{xx}^2 \rangle$ (only needed in three dimensions). Also needed for notational uniformity are the cases $m=2, n=0$ and $m=0, n=2$: for three dimensions $\langle f_x^2 \rangle = \overline{k_x^2} \langle f^2 \rangle = \frac{1}{3} \overline{k^2} \langle f^2 \rangle$ and
$\langle f_t g_{xx} \rangle = -\overline{\omega k_x^2} \langle f^2 \rangle = -\frac{1}{3} \overline{\omega k^2} \langle f^2 \rangle$ and $\langle f g_t \rangle = -\overline{\omega} \langle f^2 \rangle$, and for two dimensions
$\langle f_x^2 \rangle = \overline{k_x^2} \langle f^2 \rangle = \frac{1}{2} \overline{k^2} \langle f^2 \rangle$ and $\langle f_t g_{xx} \rangle = -\overline{\omega k_x^2} \langle f^2 \rangle = -\frac{1}{2} \overline{\omega k^2} \langle f^2 \rangle$.

There are also inequality relations: $\langle f_{xx}^2 \rangle \langle f^2 \rangle \geq \frac{9}{5} \langle f_x^2 \rangle^2$ for three dimensions and $\langle f_{xx}^2 \rangle \langle f^2 \rangle \geq \frac{3}{2} \langle f_x^2 \rangle^2$ two dimensions. Again these come from the power spectrum: for three dimensions isotropy gives $\overline{k_x^4} = \frac{1}{5} \overline{k^4} \geq \frac{1}{5} \left(\overline{k^2}\right)^2 = \frac{1}{5} \left(3\overline{k_x^2}\right)^2$, where the inequality becomes an equality if all the wavelengths are equal so that the power spectrum is a spherical shell in **k** space. For two dimensions isotropy gives $\overline{k_x^4} = \frac{3}{8} \overline{k^4} \geq \frac{3}{8} \left(\overline{k^2}\right)^2 = \frac{3}{2} \left(\overline{k_x^2}\right)^2$.

*Appendix B: Interpretation of the starting averages for 2D and 3D.*

For two dimensions, a way to understand the starting average (2) for the rate $\Omega$ of pair births and deaths per unit volume is as a simplification (using the identity $\delta(\alpha z) = |1/\alpha| \delta(z)$), of

$$\Omega = \pi \left\langle \delta(f) \, \delta(g) \, |f_y| \, |g_y| \, \delta\left(\frac{f_x}{f_y}\right) \delta\left(\frac{g_x}{g_y}\right) \left|\frac{g_t}{g_y} - \frac{f_t}{f_y}\right| \left|\frac{f_{xx}}{f_y} - \frac{g_{xx}}{g_y}\right| \right\rangle \tag{B1}$$

The interpretation of the pieces here is as follows. The first two delta functions in (3) select the zero contours of $f$ and of $g$, each being an infinitely high, sharp, curved 'ridge' following the zero contour, the product selecting those points where the ridges overlap. The two Jacobian modulus factors immediately following them give each ridge unit cross sectional integral at the places of interest where the contour is parallel to the $x$ direction. The $\delta(f_x/f_y) \, \delta(g_x/g_y)$ terms can be understood together using
$\delta(f_x/f_y) \, \delta(g_x/g_y) = \delta(g_x/g_y - f_x/f_y) \, \delta(f_x/f_y)$. The $\delta(g_x/g_y - f_x/f_y)$ requires the tangents to the two zero contours to be coincident, and the $\delta(f_x/f_y)$ is that coming from the axis fixing technique (1) for $f$, informing the calculation of the choice of $x$ axis direction (the mutual tangent direction).

The next modulus term is the relative closing speed with which the two contours approach each other immediately before the pair birth or death event. This is needed because, without it, slowly approaching curves would build up a large weight for the event, and fast ones a low weight, rather than the unit count that is required for the rate determination. Similarly the final modulus term in the numerator contains the difference in the curvatures of the two contours, or equivalently the spatial derivative of the angle difference between the two unit normal vectors to the contours. This normalizes the weight of the event so that it gives a proper unit count (otherwise, for closely matched curvatures, the delta function requiring the matching of the normal would give too high a weight).

For three dimensions the version of $\Omega_{B+D+R}$ of which (12) is the simplification is:

$$\Omega_{B+D+R} = 2\pi \left\langle \delta(f) \, \delta(g) \, |f_z| \, |g_z| \, \delta\left(\frac{f_x}{f_z}\right) \delta\left(\frac{g_x}{g_z}\right) \delta\left(\frac{f_y}{f_z}\right) \delta\left(\frac{g_y}{g_z}\right) \left|\frac{f_t}{f_z} - \frac{g_t}{g_z}\right| \times \right.$$
$$\left. \left|\left(\frac{f_{xx}}{f_z} - \frac{g_{xx}}{g_z}\right)\left(\frac{f_{yy}}{f_z} - \frac{g_{yy}}{g_z}\right) - \left(\frac{f_{xy}}{f_z} - \frac{g_{xy}}{g_z}\right)\left(\frac{f_{yx}}{f_z} - \frac{g_{yx}}{g_z}\right)\right| \right\rangle \tag{B2}$$

Here the first two delta functions with their neighbouring Jacobian modulus factors select the zero surfaces (well normalized at the points that are to be selected, where the gradient vectors are parallel, or antiparallel to the $z$ axis). The selection is done by the next four delta functions – once again the role of the two containing $g$s may be made clearer by writing $\delta(f_x/f_z) \, \delta(g_x/g_z) \, \delta(f_y/f_z) \, \delta(g_y/g_z) = \delta(f_x/f_z) \, \delta(f_y/f_z) \, \delta(g_x/g_z - f_x/f_z) \, \delta(g_y/g_z - f_y/f_z)$ the last two delta functions requiring that the two gradient vectors are parallel, or equivalently that the unit normal vectors coincide.

The next modulus term is the relative closing speed of approach of the two surfaces before or after the tangent plane contact event. This is needed to counter the weight that would otherwise obtain. The final modulus measures the matching of the two surfaces again to counter unwanted weighting. Specifically the expression inside the final modulus is the Jacobian of the mapping from the difference of the two unit normal vectors of the surfaces,to the local $x, y$ coordinates of the contact tangent plane. This is required to accompany the delta functions that make the normal vectors coincide.